\documentclass[prc,showpacs,floatfix]{revtex4}
\usepackage{bm}
\usepackage{dcolumn}
\usepackage{graphicx}
\usepackage{amsmath}
\begin{document}
\title{
{\bf Photoproduction of $\eta$ mesons on the deuteron above
$S_{11}(1535)$ in the presence of a narrow $P_{11}(1670)$
resonance}}
\author{A. Fix\footnote{Permanent address: Laboratory of mathematical physics,
Tomsk Polytechnic University, 634034 Tomsk, Russia} and L. Tiator}
\affiliation{Institut f\"ur Kernphysik, Johannes
Gutenberg-Universit\"at Mainz, D-55099 Mainz, Germany}
\author{M. V. Polyakov}
\affiliation{Institut f\"ur Theoretische Physik II,
Ruhr-Universit\"at Bochum, D-44780 Bochum, Germany}

\date{\today}

\begin{abstract}
Incoherent photoproduction of $\eta$-mesons on the deuteron is
considered. The main attention is paid to the region above the
$S_{11}(1535)$ resonance where rather narrow resonance like
structure in the total cross section extracted for $\gamma n\to\eta
n$ has been reported. The corresponding experimental results are
analyzed from the phenomenological standpoint within the model
containing a baryon $P_{11}$ with the mass about 1670 MeV and a
width less than 30~MeV. This resonance was suggested in some recent
works as a nonstrange member of the pentaquark antidecuplet with
$J^P=1/2^+$. The calculation is also performed for the polarized and
nonpolarized angular distributions of $\eta$ mesons. In addition, we
present our predictions for the cross sections of the neutral kaons
and double pion photoproduction, where the same narrow
$P_{11}(1670)$ resonance is assumed to contribute through the decay
into $K^0\Lambda$ and $\pi\Delta$ configuration.
\end{abstract}

 \maketitle

\section{Introduction}
During recent years there has been a considerable improvement of
experimental and theoretical results for the production of
$\eta$-mesons above the $S_{11}(1535)$ resonance region. Extensive
and accurate data were obtained for the reaction $\gamma p\to\eta p$
and to a lesser extent for $\gamma d\to\eta np$. Furthermore, an
amount of data is now available for $\eta$ photoproduction on
quasifree protons and neutrons, where the knocked-out nucleons are
detected in coincidence with $\eta$-mesons. The energy region of a
special interest corresponds to the excitation of the nucleon-like
resonance identified in the Particle Data Group listing as
$P_{11}(1710)$ and marked there by three stars. The first question
with respect to this particle concerns its status. This resonance
clearly seen by several partial wave analysis (PWA) groups,
e.g.~\cite{KH}, but was not found in the GWU $\pi N$ analysis
\cite{VPI,Arndt:2006bf}. Another aspect is related to the distinct
role of $P_{11}(1710)$ in the physics of pentaquarks. This state was
assumed in \cite{DPP} to be a nonstrange member of the antidecuplet
representation ($\overline{10}$) of the pentaquark multiplet
$qqqq\bar{q}$. This assumption was then used to scale the masses of
$\overline{10}$ members, and thus was the most crucial input for
further search of $\Theta^+$, the $S=+1$ member of this baryon
group. At the same time, as it is discussed in \cite{VPI2} in this
case a question arises, how to reconcile a small width of $\Theta^+$
(less than 1 MeV according to the estimations noted in \cite{VPI2})
with rather broad structure of its presumable nonstrange partner
$P_{11}(1710)$ (on average 100 MeV according to PDG~\cite{PDG}). As
is emphasized in \cite{VPI2}, this principal difference seems to
contradict an expectation that the widths of particles entering the
same multiplet should be comparable. Another problem with
$P_{11}(1710)$ as a member of $\overline{10}$ is its too large mass.
It is about 30 MeV larger than the value required by the
Gell-Mann-Okubo rule \cite{DP03} , if one takes for the mass of
$\Xi_{3/2}^{--}$, the heaviest member of $\overline{10}$, the value
1862 MeV reported in \cite{Alt}.

To resolve the discrepancy, it was suggested in
Refs.~\cite{VPI2,DP03} that another $P_{11}$ state has to be put on
the place of a nonstrange $\overline{10}$ member. Like the
$P_{11}(1710)$ this new state was assumed to be highly inelastic
with the main nonstrange decay modes $\eta N$ and $\pi\Delta$. But
contrary to $P_{11}(1710)$ it should have a smaller decay width
$\Gamma_{P_{11}}\leq 10$ MeV and a lower mass $M\approx 1670$ MeV. A
modified PWA analysis of $\pi N$ \cite{VPI2}, modified in such a way
that it becomes more sensitive to baryons with smaller widths has
shown that there are narrow regions in the baryon spectrum around
1680 and 1730 MeV where exotic states might exist. Moreover, as was
shown in \cite{PRathke}, due to the specific isotopic structure of
the $\gamma N\to P_{11}$ transition, this resonance should be
photoproduced mostly on a neutron rather than on a proton.

From the considerations above it is clear that a natural way to
verify the theory of $\overline{10}$ is to look for a narrow
structure in the energy dependence of the total cross section for
photoproduction of $\eta$ and other light mesons on neutrons. First
steps in this direction were made in the experimental works of GRAAL
\cite{Slava}, CB-ELSA \cite{Jaegle:2005,Krusche:2006}, and LNS-Tohoku
\cite{kas} where the
cross section for $\gamma n\to\eta n$ was extracted from the
reaction on the deuteron. The preliminary data really seem to
exhibit a peak around $E_{\gamma}=1$ GeV what roughly corresponds to
the $\gamma N$ invariant mass $W=1670$ MeV, if one assumes the
initial neutron to be at rest in the target.

The question about the nature of the observed peak was already
touched upon on the phenomenological basis in
\cite{Slava,Azim,Choi:2005ki}. The present paper is partially
intended to provide a calculation of the reaction $\gamma d\to\eta
np$ that would not suffer from different uncertainties, like the
oversimplified treatment of the Fermi motion and final state
interactions. Moreover we present the results for some asymmetries
of this reaction in order to study their sensitivity to the possible
contribution of a narrow $P_{11}$ state.

Another point related to the $P_{11}$ problem has a more practical
aspect. It deals with the ratio $\sigma_n/\sigma_p$ of the neutron
to the proton cross section for $\eta$ photoproduction around the
lab photon energy 1020 MeV. This ratio measured for quasifree
nucleons in \cite{Jaegle:2005,Krusche:2006} exhibits a clear
resonance behavior with the width about 100 MeV. A natural
explanation of this phenomenon would be an existence of a baryon
resonance which is predominantly excited on a neutron. In the
EtaMAID model~\cite{etaMAID} this feature is assigned to
$D_{15}(1675)$. The results of EtaMAID2001 for $\sigma_n/\sigma_p$
give quite a good account of the data
\cite{Jaegle:2005,Krusche:2006}. However the problem with
EtaMAID2001 is that the $\eta N$ decay width $\Gamma_{\eta N}\approx
0.17\, \Gamma_{tot}$ needed to fit the data for $\gamma p\to\eta p$
is an order of magnitude larger than the value, suggested by the PDG
listing (less than 1$\%$ of $\Gamma_{tot}$) and visibly exceeds the
boundaries given by the $SU(3)$ analysis (about 3$\%$ of
$\Gamma_{tot}$) \cite{Samios,Polyakov:2006}. The latter discrepancy
is especially unsatisfactory inasmuch as the $D_{15}(1675)$ state is
a member of a well established baryon octet with branchings
determined by the broken $SU(3)$ symmetry. This inconsistency was
already discussed in \cite{Krakau}. It would therefore be
interesting to see whether an inclusion of a hypothetic narrow
$P_{11}(1670)$ baryon into the production amplitude instead of a too
strong $D_{15}(1675)$ could provide a reasonable explanation of the
structure seen in $\sigma_n/\sigma_p$. Although the nucleon-like
member of $\overline{10}$ is assumed to be very narrow, the effect
of broadening needed to fit the data might be due to Fermi motion in
the deuteron.

In addition to the previous works where only the $\eta N$ channel
was explored, we also pay some attention to kaon and double pion
photoproduction. According to the results of \cite{VPI2} there is a
non-vanishing probability for the exotic $P_{11}$ to decay into
$K\Lambda$ and $\pi \Delta$ states. For instance, the $\pi\Delta$
mode was estimated in \cite{VPI2} on the level of 30\,$\%$. If the
members of $\overline{10}$ are predominantly excited on a neutron,
it is clear that from the experimental point of view, the easiest
way to investigate the $P_{11}\to\pi\Delta$ transition is to measure
the reaction $\gamma d\to\pi^-\pi^0pp$. The other channels will
contain strong proton contributions, and like in the $\eta$ case
this will lead to the necessity to single out the neutron reaction.
So far in the region around $E_\gamma=1$ GeV only one measurement of
$\gamma d\to\pi^-\pi^0pp$ was reported in \cite{Shiffer}, which
furthermore has very large experimental uncertainties. Here we
present our results for the corresponding total cross sections,
where the role of $P_{11}(1670)$ in $\pi\pi$ and kaon
photoproduction can be seen.

In section II we outline the model used for our calculation of
$\gamma d\to\eta np$ and show our results for different channels and
observables in section III. In section IV we close with a summary
and conclusions.

\section{Formalism}\label{formal}

In this section we briefly review the formal ingredients of the
quasifree $\eta$ photoproduction on the deuteron
\begin{equation}\label{0}
\gamma+d\to\eta+p+n\,.
\end{equation}
Taking as independent variables the 3-momentum of the proton
$\vec{p}_p$ in the deuteron lab frame and the spherical angle
$\Omega_{q^*}$ of the $\eta$-meson momentum $\vec{q}\,^*$ in the
$\eta n$ c.m.\ system, one obtains for the unpolarized cross section
\begin{equation}\label{10}
\frac{d\sigma}{d\vec{p}_p\,d\Omega_{q^*}}=\frac{1}{6}\ {\cal
K}\!\sum\limits_{sm_s\lambda m_d} |\langle
sm_s|\,\hat{T}\,|\lambda\, 1m_d\rangle|^2\,,
\end{equation}
where the phase space factor reads
\begin{equation}\label{20}
{\cal K}=\frac{1}{(2\pi)^5}
\frac{M_N^2}{4\omega_{\gamma}}\frac{|\vec{q}\,^*|}{E_pW_{\eta n}}\,.
\end{equation}
In Eq.~(\ref{20}) $\omega_{\gamma}$ and $E_p$ are the photon and the
proton energies in the lab frame and $M_N$ and $W_{\eta n}$ are the
nucleon mass and the $\eta n$ invariant energy. The final state is
determined by total spin $(sm_s)$ of the $NN$ system. The index
$\lambda$ denotes the photon polarization and $(1m_d)$ stands for
the deuteron spin.

As for the polarization observables, we considered only the
asymmetries having counterparts in the elementary reaction $\gamma
N\to\eta N$. Namely, wee took the (linear) beam asymmetry $\Sigma$,
the target asymmetry for vector polarized deuterons $T^0_{11}$, and
the beam-target asymmetry $T^c_{10}$ for circularly polarized
photons and vector polarized deuterons. For their definitions we
take the formulas similar to those obtained in Ref.~\cite{ArFi}
\begin{eqnarray}
\Sigma\frac{d\sigma}{d\Omega_{q^*}}&=&\frac{1}{3}\int
d\vec{p}_p\,{\cal K}\sum\limits_{sm_sm_d}
T^*_{sm_s1m_d}T_{sm_s-1m_d}\label{51}\,,\\
T^0_{11}\frac{d\sigma}{d\Omega_{q^*}}&=&\frac{2}{3}\int
d\vec{p}_p\,{\cal K}\sum\limits_{sm_sm_d}
\left(T^*_{sm_s1-1}T_{sm_s10}+T^*_{sm_s10}T_{sm_s11}\right)\label{52}\,,\\
T^c_{10}\frac{d\sigma}{d\Omega_{q^*}}&=&\int d\vec{p}_p\,{\cal
K}\sum\limits_{sm_sm_d}
\left(|T_{sm_s11}|^2-|T_{sm_s1-1}|^2\right)\label{53}
\end{eqnarray}
with $T_{sm_s\lambda m_d}=\langle sm_s|\,\hat{T}\,|\lambda\,
1m_d\rangle$. Note that the last observable (\ref{53}) differs by a
factor $\frac{1}{\sqrt{6}}$ from that given in \cite{ArFi}. In this
form it is equal to the familiar GDH integrand for the deuteron
\cite{ArFiSch}. In a simpler form the corresponding formulas read
\begin{equation}\label{55}
\Sigma\frac{d\sigma}{d\Omega_{q^*}}=\frac{1}{2}
\frac{d\sigma_{\perp}-d\sigma_{\parallel}}{d\Omega_{q^*}}\,.
\end{equation}
Here $\sigma_{\perp}$ ($\sigma_{\parallel}$) is the cross section
with the photon beam polarized perpendicular (parallel) to the
reaction plane.
\begin{equation}\label{60}
T^0_{11}\frac{d\sigma}{d\Omega_{q^*}}=\frac{1}{2}
\frac{d\sigma_+-d\sigma_-}{d\Omega_{q^*}}\,,
\end{equation}
where $\sigma_+$\,$(\sigma_-)$ corresponds to the target polarized
"up" ("down") in the direction of $\vec{k}\times\vec{q}$. Finally,
the spin asymmetry (\ref{53}) reads
\begin{equation}\label{65}
T_{10}^c\frac{d\sigma}{d\Omega_{q^*}}=\frac{d\sigma^P-d\sigma^A}{d\Omega_{q^*}}\,,
\end{equation}
where $d\sigma^P$ and $d\sigma^A$ correspond to the photon spin
oriented parallel and antiparallel to the deuteron spin. In the case
of a free nucleon, the vector polarization $T^0_{11}$ corresponds to
the target polarization $T$ and the beam-target double polarization
$T_{10}^c$ corresponds to the polarization observable $E$. More
precisely, if we neglect Fermi motion we get
\begin{equation}
T^0_{11} \rightarrow -T \quad \mbox{and} \quad T_{10}^c \rightarrow
-E\,.
\end{equation}
The effect of Fermi motion can best be seen on the fact that in
parallel kinematics the absolute value of the double polarization
$T_{10}^c$ is not exactly 1 as in the case of a free nucleon.

To construct the single nucleon operator for $\eta$ photoproduction
we use the $\eta$-MAID analysis developed in \cite{etaMAID} and
\cite{etaRegge}. The resonance contribution in the partial wave
$\alpha$ is parameterized in terms of standard Breit-Wigner
functions
\begin{equation} \label{62}
t_{\gamma,\eta}^\alpha(R\,; \lambda) = \tilde{A}_{\lambda}\,
\frac{\Gamma_{tot}\,W_R}{W_R^2-W^2-iW_R\Gamma_{tot}}\, f_{\eta
N}(W)\,\zeta_{\eta N}\,,
\end{equation}
where a hadronic phase $\zeta_{\eta N}=\pm1$ fixes a relative sign
between the $N^* \rightarrow \eta N$ and the $N^* \rightarrow \pi N$
couplings. For most of the states the phases $\zeta_{\eta N}$ were
treated as free parameters in the fitting procedure. As principal
fit parameters of our analysis we used the resonance masses $W_R$,
the total widths $\Gamma_R=\Gamma_{tot}(W_R)$, the branching ratios
$\beta_{\eta N}=\Gamma_{\eta N}(W_R)/\Gamma_R$ and the photon
couplings $\tilde{A}_{\lambda}=\{A_{1/2},A_{3/2}\}$. However, we fix
those parameters, where reliable results are given by PDG, see
Tables~\ref{ta1} and~\ref{ta2}. For more details of the
parametrization of the energy-dependent widths and the vertex
function $f_{\eta N}(W)$ we would like to refer the reader to
Refs.~\cite{etaMAID} and \cite{etaRegge}.

\begin{table}[htb]
\renewcommand{\arraystretch}{1.2}
\caption{Parameters of nucleon resonances from EtaMaid2001
\protect\cite{etaMAID} with standard vector meson poles, model (I).
The masses and widths are given in MeV, $\beta_{\eta N}$ is the
branching ratio for the eta decay channel and $\zeta_{\eta N}$ the
relative sign between the $N^* \rightarrow \eta N$ and the $N^*
\rightarrow \pi N$ couplings. The photon couplings to the proton and
neutron target for helicity $\lambda=$1/2 and 3/2 are given in units
of $10^{-3}/\sqrt{\mbox{GeV}}$. The underlined parameters are fixed
and are taken from PDG~\protect\cite{PDG}. The asterisk for
$_nA_{1/2}$ of the $S_{11}(1535)$ denotes a fixed $n/p$ ratio
obtained from the threshold experiment\protect\cite{Krusche0}.}
\begin{ruledtabular}
{\begin{tabular}{ccccccccc}
 $N^*$ & Mass & Width & $\beta_{\eta N}$ & $\zeta_{\eta N}$ & $_pA_{1/2}$ & $_pA_{3/2}$ & $_nA_{1/2}$ & $_nA_{3/2}$ \\
\hline
 $D_{13}(1520)$ & \underline{1520}  & \underline{120} & $0.06\%$ & $+1$ & $-52$ & \underline{166} & $\underline{-41}$ & $\underline{-135}$ \\
 $S_{11}(1535)$ & 1541  & 191 & $  \underline{50\%}$ & $+1$ & 118 & $-$ & $-97^*$  & $-$ \\
 $S_{11}(1650)$ & 1638  & 114 & $  8\%$ & $-1$ &  68 & $-$ & $-56$ & $-$ \\
 $D_{15}(1675)$ & 1665  & \underline{150} & $ 17\%$ & $-1$ &  18 &  24 & $\underline{-43}$ & $\underline{-58}$ \\
 $F_{15}(1680)$ & 1682  & \underline{130} & $0.06\%$ & $+1$ & $-21$  & 124 &  \underline{52} &  $\underline{-41}$ \\
 $D_{13}(1700)$ & \underline{1700}  & \underline{100} & $0.3\%$ & $-1$ & $\underline{-18}$ &  $\underline{-2}$ &  \underline{0}  & $\underline{-3}$ \\
 $P_{11}(1710)$ & 1720  & \underline{100} & $  26\%$ & $+1$ &  23 & $-$ & $\underline{-2}$ & $-$ \\
 $P_{13}(1720)$ & \underline{1720}  & \underline{150} & $ 3\%$ & $-1$ &  \underline{18} & $\underline{-19}$ & $\underline{1}$ & $\underline{-29}$ \\
\end{tabular} \label{ta1}}
\end{ruledtabular}
\end{table}
\begin{table}[htb]
\renewcommand{\arraystretch}{1.2}
\caption{Parameters of nucleon resonances from EtaMaid2003
\protect\cite{etaRegge} with reggeized vector mesons, model (II).
Notation as in Table~\protect\ref{ta1}.}
\begin{ruledtabular}
\begin{tabular}{ccccccccc}
$N^*$ & Mass & Width & $\beta_{\eta N}$ & $\zeta_{\eta N}$ & $_pA_{1/2}$ & $_pA_{3/2}$ & $_nA_{1/2}$ & $_nA_{3/2}$ \\
\hline
 $D_{13}(1520)$ & \underline{1520}  & \underline{120} & $0.04\%$ & $+1$ & $\underline{-24}$ & \underline{166} & $\underline{-59}$ & $\underline{-139}$ \\
 $S_{11}(1535)$ & 1521  & 118 & $  \underline{50\%}$ & $+1$ &  80 &  $-$  &  $-65^*$  &   $-$  \\
 $S_{11}(1650)$ & 1635  & 120 & $  16\%$ & $-1$ &  \underline{46} &  $-$  & $\underline{-38}$ & $-$ \\
 $D_{15}(1675)$ & 1665  & \underline{150} & $ 0.7\%$ & $+1$ &  19 &  15 & $\underline{-43}$ & $\underline{-58}$ \\
 $F_{15}(1680)$ & 1670  & \underline{130} & $0.003\%$ & $+1$ & $-15$ & 133 & \underline{29} & $\underline{-33}$ \\
 $D_{13}(1700)$ & \underline{1700}  & \underline{100} & $0.025\%$ & $-1$ & $\underline{-18}$ & $\underline{-2}$ & \underline{0} & $\underline{-3}$ \\
 $P_{11}(1710)$ & 1700  & \underline{100} & $26\%$ & $-1$ & \underline{9} &  $-$ & $\underline{-2}$ & $-$ \\
 $P_{13}(1720)$ & \underline{1720} & \underline{150} & $4\%$ & $+1$ &  \underline{18} & $\underline{-19}$ & \underline{1} &  $\underline{-29}$ \\
\end{tabular}\label{ta2}
\end{ruledtabular}
\end{table}

As mentioned above, the standard EtaMAID model \cite{etaMAID}
contains a too large contribution of $D_{15}(1675)$ which is
inconsistent with rather small mean value of $\beta_{\eta N}$ given
by PDG as well as with the $SU(3)$ symmetry bounds
\cite{Polyakov:2006}. An alternative approach to $\gamma N\to\eta N$
was implemented in \cite{etaRegge} where the experimentally observed
increase of $\eta$ production on the proton in the forward direction
is reproduced by Regge poles in the $t$ channel. In this case the
contribution of $D_{15}$ can be weakened down to 0.7$\%$ in
accordance to the PDG analysis. At the same time, as will be
discussed in the next section, this model being fitted to the proton
data cannot account for the preliminary results of $\gamma n\to\eta
n$, where the calculated cross section in the region $E_\gamma>900$
MeV exhibit a smooth energy dependence, whereas the data show a
pronounced peak.

In the present paper we consider two models for $\gamma N\to\eta N$:
(i) the standard EtaMAID model \cite{etaMAID}, with strong coupling
of $D_{15}(1675)$ to the $\eta N$ state ($\Gamma_{\eta
N}=0.17\,\Gamma_{tot}$), and (ii) the modified reggeized MAID model
\cite{etaRegge} in which an additional narrow $P_{11}$ state is
inserted. The parameters of the $P_{11}$ resonance are listed in
Table~\ref{ta3}.
\begin{table}[htb]
\renewcommand{\arraystretch}{1.2}
\caption{Mass, total width, $\eta N$ branching ratio and photon
helicity couplings in units of $(10^{-3}/\sqrt{\mbox{GeV}})$ for the
$P_{11}$ pentaquark state in our calculation.}
\begin{ruledtabular}
\begin{tabular}{ccccccc}
Mass & Width & $\beta_{\eta N}$ & $\beta_{\pi\Delta}$ &
$\beta_{K\Lambda}$ & $_pA_{1/2}$ & $_nA_{1/2}$ \\
\hline
1670 & 10(30) & 40\% & 30\% & 30\% & 8 & 30(35) \\
\end{tabular}
\end{ruledtabular}
\label{ta3}
\end{table}
For the total width we use two values $\Gamma_{tot}(P_{11})=10$ MeV
and $\Gamma_{tot}(P_{11})=30$ MeV. The former was found in
Ref.~\cite{VPI2} taking the octet-antidecuplet mixing angle from
\cite{DPP}. As is noted in \cite{VPI2} more recent information about
this parameter points to a larger value of
$\Gamma_{\pi\Delta}(P_{11})$ which might result in an increasing
total $P_{11}$ width up to $\sim$ 30 MeV. Furthermore, as pointed
out in \cite{VPI2} this latter value should be an upper bound of the
resonance width treated as undetectable in their modified $\pi N$
PWA. The neutron photocoupling of $P_{11}(1670)$ is chosen according
to the phenomenological analysis of Ref.~\cite{Azim} and the chiral
quark-soliton model calculations of Ref.~\cite{Yang}.

To calculate the cross section on the deuteron we use a production
operator valid in an arbitrary frame of reference. For this purpose
we resort to the standard scheme to pass from the CGLN amplitudes
$F_i$ to the Lorentz invariant amplitudes $A_i$ (see, e.g.,
\cite{BeTa,Pasquini:2006}). The effects of final state interaction
(FSI) were included within a method where rescatterings within the
pairs of particles are taken into account up to the first order in
the corresponding two-body $t$-matrices. In general we have
\begin{equation}\label{75}
\hat{T}=\hat{T}^{IA}+\hat{T}^{NN}+\hat{T}^{\eta N}\,,
\end{equation}
where the first term on the r.h.s.\ stands for the impulse
approximation, which for the noncoherent reaction is equivalent to
the spectator nucleon model. The terms $T^{NN}$ and $T^{\eta N}$
include one-loop $NN$ and $\eta N$ rescattering mechanisms as is
explained in \cite{FiAr97}. The operator $\hat{T}^{IA}$ can
schematically be presented as a sum of the elementary operators
referring to a proton and a neutron
\begin{equation}\label{77}
\hat{T}^{IA}=\hat{t}_p(1)+\hat{t}_n(2)\,,
\end{equation}
where $\hat{t}(i)$ acts on the $i$th nucleon. The actual results,
e.g. for a quasifree neutron, are obtained by setting $\hat{t}_p=0$
in the expression (\ref{77}). The last two terms in Eq.~(\ref{75})
contain interaction between the corresponding particles. For formal
details, concerning the treatment of FSI effects in $\gamma d\to\eta
np$ we refer the reader to Ref.~\cite{FiAr97}.

\section{Results and discussion}\label{results}

Before going to the details it is instructive to estimate possible
contributions of $P_{11}(1670)$ into different channels. For this
purpose we use the formula which gives the partial cross section at
the resonance position $W=W_R$
\begin{equation}\label{80}
\sigma_{X}=\xi_X(2J+1)\frac{\pi}{\omega_\gamma^2}\frac{\Gamma_{\gamma
N^*}\Gamma_{N^*\to X}}{\Gamma_{tot}^2}\,,\quad X\in\{\eta n,\,\pi\pi
n,\, K^0\Lambda\}\,.
\end{equation}
Here the factor $\xi_X$ $(0<\xi_X\leq 1)$ takes into account
isotopic separation of the $N^*\to X$ vertex. In particular one
obtains
\begin{equation}\label{85}
\xi_{\eta n}=\xi_{K^0\Lambda}=1\,,\quad \xi_{\pi^+\pi^-
n}=5/9\,,\quad \xi_{\pi^-\pi^0 n}=\xi_{\pi^0\pi^0n}=2/9\,.
\end{equation}
Taking $J=1/2$ and the $P_{11}(1670)$ helicity amplitudes as well as
the widths from Table~\ref{ta3} (for a moment we take
$\Gamma_{tot}(P_{11})=10$~MeV) we will have at the resonance point
$W=W_R$
\begin{equation}\label{90}
\sigma_{\eta n}\approx 15\,\mu b\,,\quad \sigma_{\pi^+\pi^-
n}\approx 6.7\,\mu b\,,\quad \sigma_{\pi^-\pi^0
n}=\sigma_{\pi^0\pi^0 n}\approx 2.7\,\mu b\,,\quad
\sigma_{K^0\Lambda}\approx 11\,\mu b\,.
\end{equation}
The so obtained $P_{11}$ partial cross sections have to be compared
to the characteristic values of the corresponding total cross
sections in the region $E_\gamma\approx 1020$ MeV
\begin{equation}\label{95}
\sigma_{\eta n}\approx 13\,\mu
b\,\mbox{\protect\cite{etaMAID}}\,,\quad \sigma_{K^0\Lambda}\approx
2.5\,\mu b\,\mbox{\protect\cite{Mart}}\,,
\end{equation}
\begin{equation}\label{100}
\sigma_{\pi^+\pi^- n}\approx 70\,\mu
b\,\mbox{\protect\cite{2pi}}\,,\quad \sigma_{\pi^-\pi^0 n}\approx
40\,\mu b\,\mbox{\protect\cite{2pi}}\,,\quad \sigma_{\pi^0\pi^0
n}\approx 4\,\mu b\,\mbox{\protect\cite{2pi}}\,.
\end{equation}

Thus, already these simple estimations show that the channel $\eta
n$ provides the largest value of the ratio ($\sigma_X$/background)
and might be the most favorable. In other channels the signal from
$P_{11}$ is expected to be quite weak.

\begin{figure}[htb]
\includegraphics[scale=0.9]{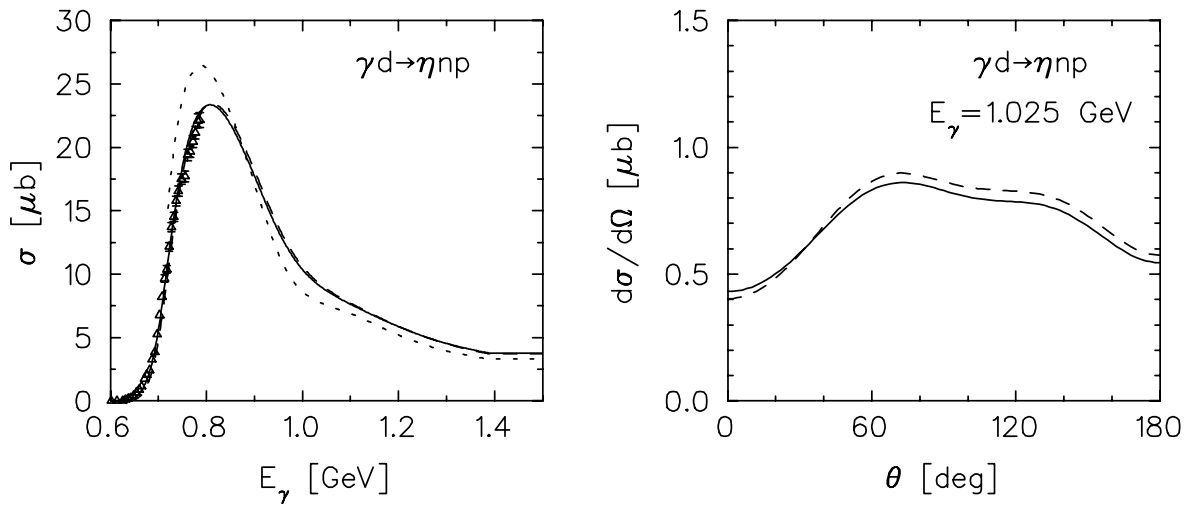}
\caption{Left panel: total cross section for $\gamma d\to\eta np$.
Right panel: $\eta$ angular distribution for $\gamma d\to\eta np$
calculated in the $\eta n$ c.m.\ frame. The dashed curves show the
results of the impulse approximation and the solid curves include
interaction between the final particles. The dotted curve on the
left panel is the sum of the total cross sections on the free proton
and neutron. For the elementary amplitude the EtaMaid2001
model~\protect\cite{etaMAID} is used. The data points are from
Ref.~\protect\cite{Krusche0}.}\label{fig1}
\end{figure}

We continue our discussion turning to the total cross section of
$\gamma d\to\eta np$ shown in Fig.~\ref{fig1} as a function of the
photon laboratory energy. Firstly, it demonstrates the influence of
the final state interaction (FSI) which is visualized as a
scattering of the produced particles in the two-body subsystems $NN$
and $\eta N$. As is explained in \cite{FiAr97} the effect is
essential in the region $E_\gamma<700$ MeV, so that its inclusion is
absolutely necessary to restore the agreement with the low-energy
data \cite{Krusche0}. In the vicinity of the $S_{11}(1535)$ peak the
FSI effect mostly comes from the absorbtion of the produced $\eta$
mesons on the deuteron through the transition to pions. This
situation is quite typical for reactions proceeding through the
resonance excitation. For heavier nuclei it results in a strong
absorption rate \cite{Landau}, so that the interaction of the
produced $\eta$-mesons with the target is predominantly of a
diffractive character. Clearly, in our case this effect is softened
by a small number of nucleons in the deuteron. With increasing
energy the influence of FSI vanishes, so that the IA should be
considered as a good approximation to the reaction dynamics. This
result is quite foreseeable, since both $NN$ and $\eta N$
interaction are important almost exclusively in the $s$-waves which
occupy only a very small fraction of the phase space available at
high energies. The influence of FSI becomes more visible in the
$\eta$ angular distribution presented on the right panel of
Fig.~\ref{fig1}. However, the general validity of IA shows that our
notion about the process on the deuteron as the one taking place on
one of the nucleons with another nucleon mainly behaving as a
spectator is justified.

\begin{figure}[htb]
\includegraphics[scale=.9]{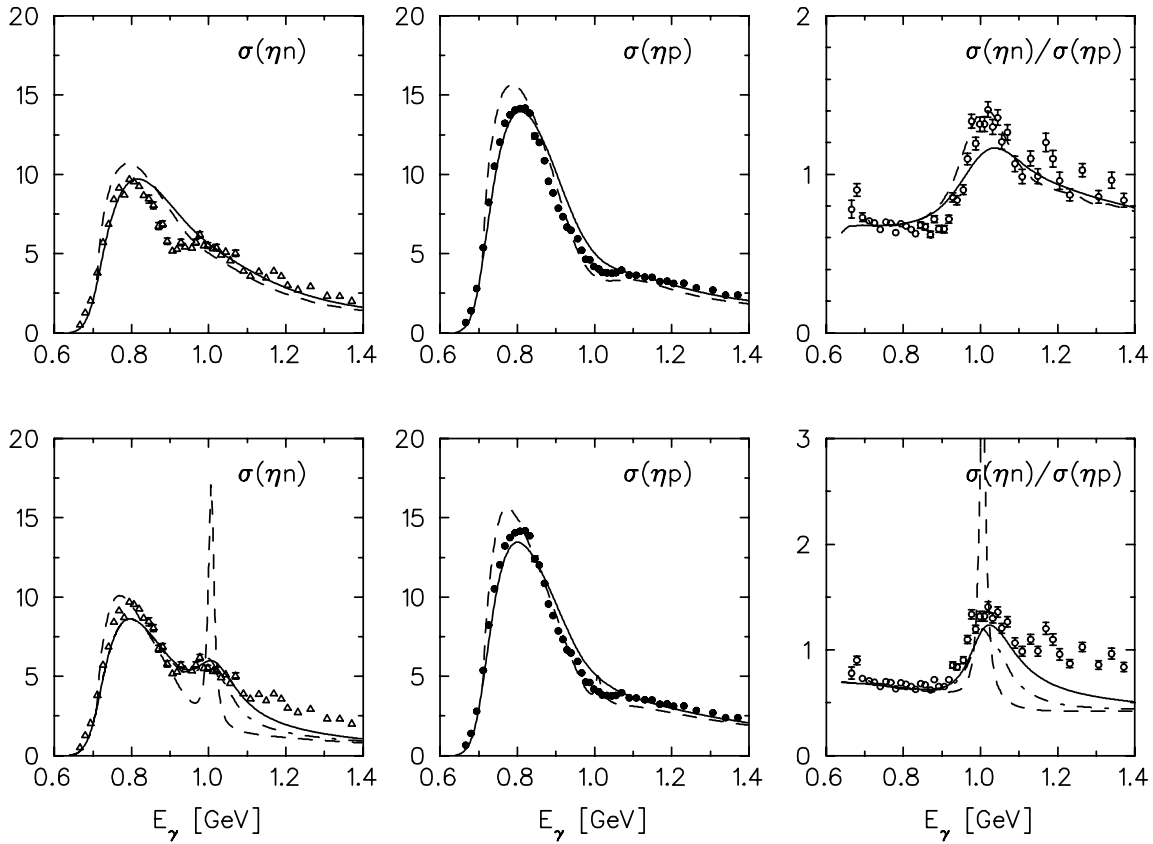}
\caption{Total cross section for noncoherent $\eta$ photoproduction
on a deuteron versus the photon lab energy. The results are
presented for the strong $D_{15}$ model (i) (upper row) and the
narrow $P_{11}$ model (ii) (lower row). The dashed curves show the
corresponding single nucleon cross sections. The dash-dotted and the
solid curves on the left and the right panels in the lower row are
obtained with $\Gamma_{P_{11}}=10$ MeV and $\Gamma_{P_{11}}=30$ MeV
respectively. The free nucleon cross section corresponds to
$\Gamma_{P_{11}}=10$ MeV. The preliminary data from
Ref.~\cite{Krusche:2006} are normalized to the EtaMAID calculations
in the first maximum.} \label{fig2}
\end{figure}

In Fig.~\ref{fig2} we show our results for the total cross section
on quasifree nucleons calculated within the strong $D_{15}$ approach
(i) and the narrow $P_{11}$ model (ii). In the same figure the
dashed lines are the corresponding single nucleon cross sections.
Firstly, as we can see, the Fermi motion strongly influences the
parameters of the $P_{11}$ peak. Namely its width becomes about 7
times larger in comparison to the free nucleon, whereas its position
remains unchanged. In this context we also note that even though the
peak observed in \cite{Jaegle:2005,Krusche:2006} is quite narrow,
the width $\Gamma_{tot}(P_{11})=10$ MeV seems to be somewhat too
small in order to fit the experimental results. If we change from 10
MeV to 30 MeV the agreement with the CB-ELSA data is improved.

It is interesting that although the model (i) does not reproduce the
shape of the quasifree neutron cross section around $E_\gamma\approx
0.9$ GeV (left upper panel in Fig.~\ref{fig2}), it gives quite a
good description of the rapidly varying quantity $\sigma_n/\sigma_p$
presented on the right panel. Analyzing the energy dependence of
$\sigma_n$ and $\sigma_p$ one can conclude that a narrow structure
in this ratio is primarily due to a shoulder in the proton cross
section. It becomes less distinct in the deuteron, because of
filling the slight minimum in $\sigma_p$ at $E_\gamma=1$ GeV, what
should be a natural consequence of a smearing effect caused by the
Fermi motion. As a result, the theoretical ratio $\sigma_n/\sigma_p$
turns down from 1.4 to about 1.2. At the same time, it is somewhat
strange that this natural change is not supported by the preliminary
experimental results~\cite{Krusche:2006} (filled circles in the
middle panels). The latter remain to be better described by the MAID
calculation for a free proton.

As is shown in Fig.~\ref{fig1} the model predicts a visible
modification of the form and the height of the $S_{11}(1535)$ peak
when we pass from the free nucleon to the bound nucleon in the
deuteron. The main reason is that the phase space factor $\cal K$
for three particles (\ref{20}) increases more slowly than the
two-body phase space in the elementary process. As a result, the
resonance form of the amplitude folded by $\cal K$ is slightly
shifted to the right on the energy scale. In the vicinity of
$P_{11}(1670)$, where the effects caused by the mass difference
between the nucleon and the deuteron become insignificant, the peak
position in the reaction on a deuteron nearly coincide with the one
on the free nucleon.

\begin{figure}[htb]
\includegraphics[scale=.7,angle=90]{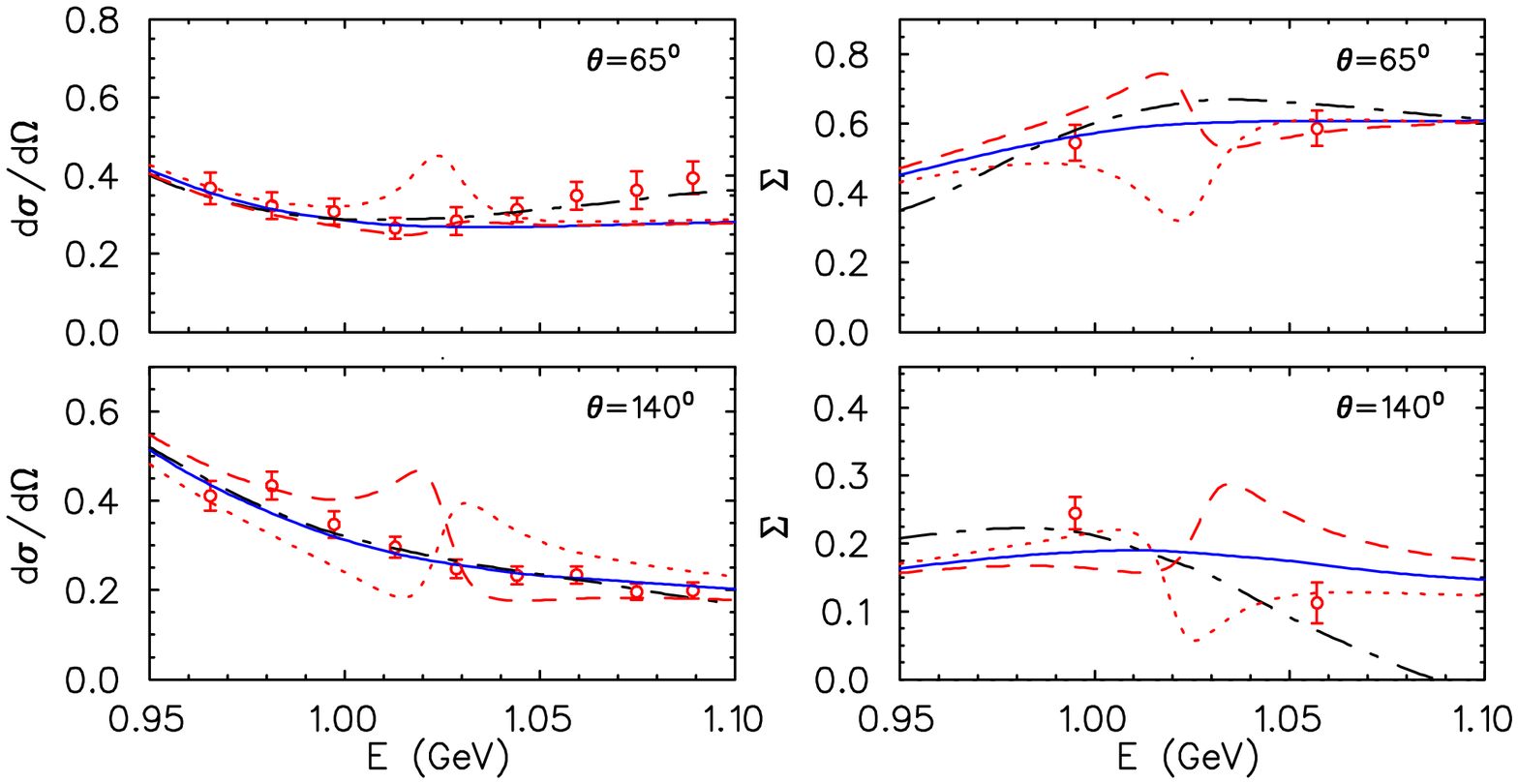}
\caption{Differential cross section and beam asymmetry for $\eta$
photoproduction on a free proton. The solid and the dash-dotted
curves are calculated with the Reggeized
model~\protect\cite{etaRegge} (without $P_{11}(1670)$) and the
ordinary EtaMAID model (i)~\protect\cite{etaMAID}. Addition of the
narrow $P_{11}$ resonance with $\zeta_{\eta N}=+1(-1)$ into the
Reggeized amplitude gives the dashed (dotted) curve. The data are
from GRAAL, Ref.~\protect\cite{GRAAL2} for the cross section and
Ref.~\protect\cite{GRAAL2001} for the beam asymmetry.}\label{fig2a}
\end{figure}

As we can see from Fig.~\ref{fig2} our calculation with
$A_{1/2}^p\approx\frac{1}{3}A_{1/2}^n$ predicts quite a pronounced
peak also in the total cross section on a free proton. In this
connection it seems to be reasonable to analyze the available data
for $\gamma p\to\eta p$ \cite{GRAAL2,CLAS,Crede,Naka} using finer
energy binning in the corresponding energy region. As an example we
plotted in Fig.~\ref{fig2a} our predictions for the
$d\sigma/d\Omega_{q^*}$ and $\Sigma$ asymmetry for three different
$\eta$ angles in $\eta N$ c.m.\ system. In particular, the results
show that the nonpolarized cross section is more sensitive to the
$P_{11}$ contribution in the backward angular region, whereas in the
beam asymmetry this resonance should be visible in both hemispheres.

\begin{figure}[htb]
\includegraphics[scale=.7]{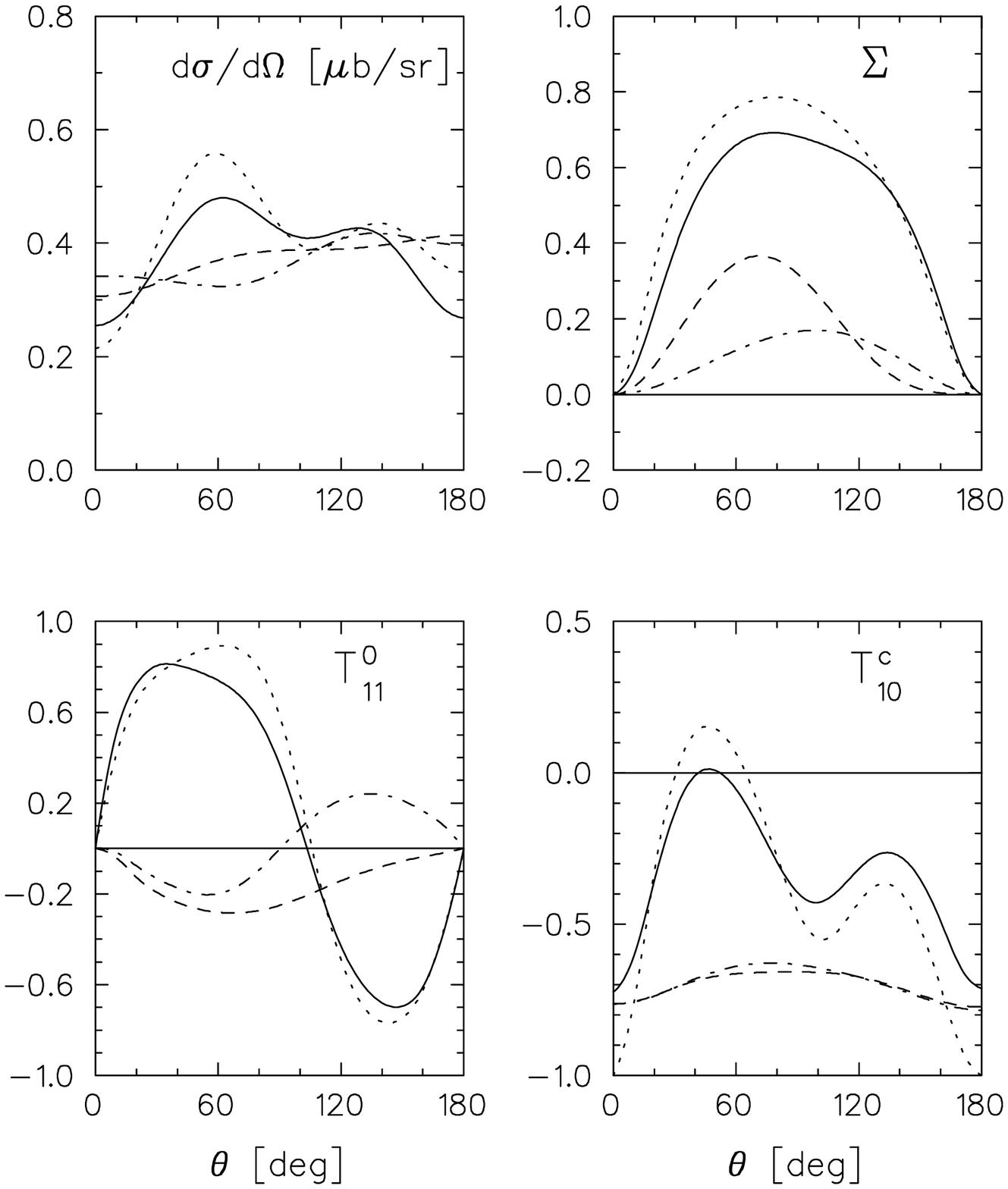}
\caption{Unpolarized differential cross section and spin asymmetries
for $\eta$ photoproduction on a quasifree neutron in a deuteron
calculated at $E_\gamma=1020$~MeV. The solid curves show the results
of EtaMAID2001 with standard vector meson poles and a strong
$D_{15}(1675)$ resonance. The dashed (dash-dotted) curevs show the
results with the reggeized model and a narrow $P_{11}(1670)$
resonance with a hadronic phase $\zeta_{\eta N}=+1(-1)$. The dotted
lines show the observables on a free neutron, where
$T^0_{11}\rightarrow -T$ and $T^c_{10}\rightarrow -E$. The
asymmetries are defined in Eqs.~(\ref{51}) to
(\ref{53}).}\label{fig3}
\end{figure}

As already noted, while both reaction mechanisms (i) and (ii) give
similar results for the total cross section, due to the different
orbital momentum of the $D_{15}$ and $P_{11}$ resonances, they will
show up with different angular distributions. This difference is
clearly seen in Fig.~\ref{fig3}, where we show our calculations for
a neutron at $E_\gamma=1020$~MeV for a) the strong $D_{15}$ model
(i), b) the narrow $P_{11}$ model (ii) with a phase $\zeta_{\eta
N}=+1$ and c) the narrow $P_{11}$ model (ii) with a phase
$\zeta_{\eta N}=-1$.
In the angular distribution the hadronic phase becomes important,
since the $P_{11}$ partial wave interferes with other partial waves
like the $S_{11}$. In the total cross section it can only interfere
with other contributions in the same partial wave, e.g. from the
background, and the difference can hardly be seen. As we can see the
model (ii) predicts quite a weak linear dependence of the
unpolarized cross section, what is obviously explained by the
dominance of the electric $E_{0+}$ and magnetic $M_{1-}$ dipole
amplitudes. In the ideal case of only $S_{11}$ and $P_{11}$ partial
waves one obtains
\begin{equation}\label{110}
d\sigma/d\Omega\sim|E_{0^+}|^2+|M_{1^-}|^2-2{\cal
R}e(E_{0^+}^*M_{1^-})\cos\theta\,.
\end{equation}
For $S_{11}$ and $D_{15}$ we have a more complicated angular
dependence which is symmetric with respect to $\cos\theta=0$
\begin{equation}\label{115}
d\sigma/d\Omega\sim A+B\cos^2\theta+C\cos^4\theta\,,
\end{equation}
where the coefficients $A,B$, and $C$ are quadratic forms of the
multipoles $E_{0^+}$, $E_{2^+}$, and $M_{2^+}$. Of course, in the
real situation the observed cross section can not be explained
entirely by only two resonances so that this ideal picture is
distorted by the presence of other, less important, states with
different quantum numbers.

In the same figure we demonstrate angular distributions of
$\eta$-mesons, obtained with polarized photon beam and target. As
already noted, for the polarization observables we took only those,
which can easily be interpreted for quasifree reactions. The photon
asymmetry $\Sigma$ given by the strong $D_{15}$ model (i) visibly
overestimates that provided by the narrow $P_{11}$ model. In the
extreme case of pure $D_{15}$ amplitude we will have
$\Sigma(\theta=90^o)=1$. On the contrary, in the model (ii) the
nontrivial $\Sigma$ asymmetry is only due to interference of the
$S_{11}$ and the $P_{11}$ waves with higher spin states. The same is
true for the vector target asymmetry $T_{11}^0$. Here again the
model (i) demonstrates typical $D_{15}$ behavior with strong rise at
forward angles and crossing zero close to $\theta=90^o$. The model
(ii) gives the nonzero value of $T_{11}^0$ due to interference of
the dominating resonances $S_{11}$ and $P_{11}$ with other resonance
states having $J>1/2$. The last asymmetry is therefore relatively
small and should be more model dependent. The angular distribution
shown by the GDH-asymmetry $T_{10}^c$ is what one could expect from
the results for the nonpolarized cross section on the left upper
panel. Here the model (ii) with the dominating $S_{11}+P_{11}$ mode
contributes only to the antiparallel component $d\sigma^A$ in
$(\ref{53})$. As a consequence, one sees about the doubled value of
the corresponding nonpolarized cross section in accordance with the
definition (\ref{53}), which again exhibits quite a weak angular
dependence. The $D_{15}$ resonance gives a strong parallel component
$d\sigma^P$, so that the resulting asymmetry qualitatively
reproduces the unpolarized angular distribution with two distinct
peaks.

\begin{figure}[htb]
\includegraphics[scale=.8]{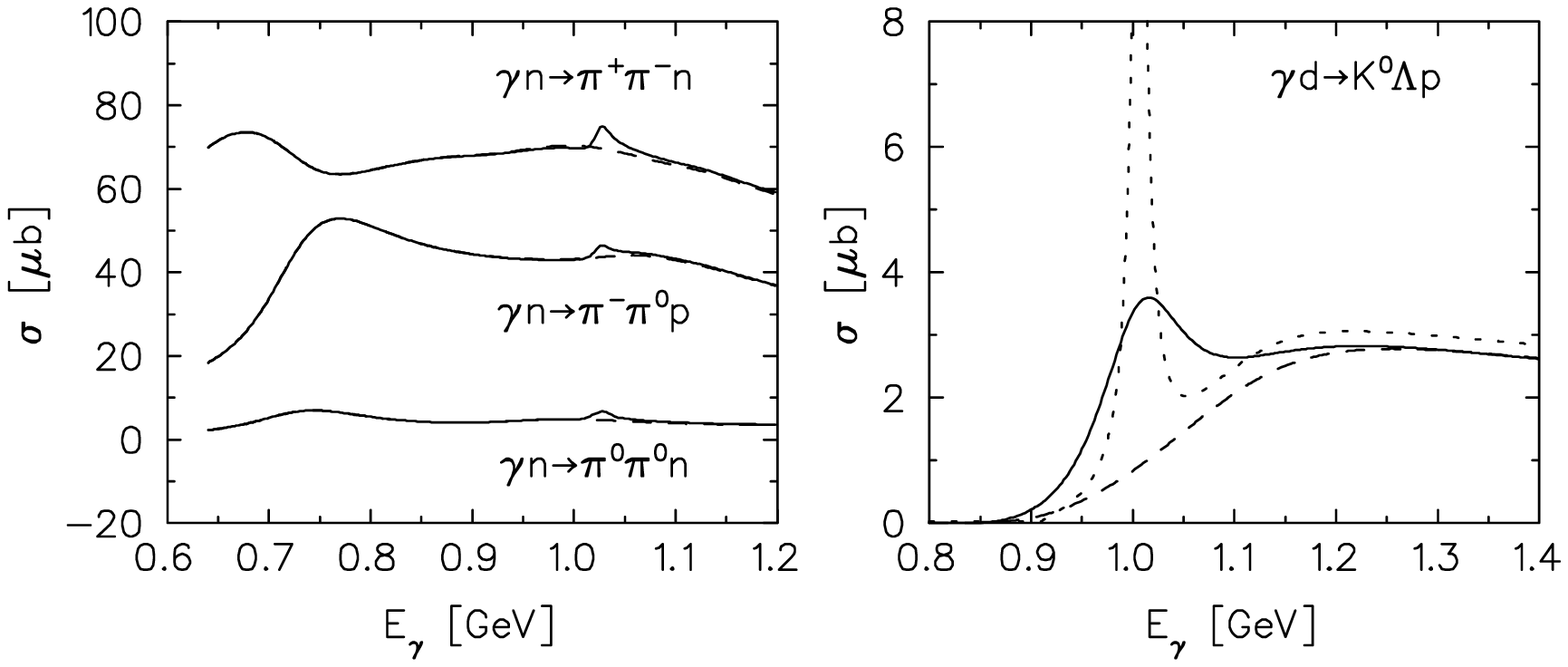}
\caption{Total cross section for $\gamma n\to\pi\pi N$ and $\gamma
d\to K^0\Lambda p$. Dashed and solid curves are obtained with and
without a narrow $P_{11}(1670)$ resonance. The dash-dotted curve on
the right panel is the corresponding free neutron cross section.}
\label{fig5}
\end{figure}

As for the other channels $\pi\pi$ and $K\Lambda$, here the
situation is less clear. According to our estimates at the beginning
of this section, these channels seem to be not so effective as a
tool to detect $P_{11}(1670)$. As an example, in Fig.~\ref{fig5} we
present our predictions for the corresponding cross sections. In the
$\pi\pi$ case the corresponding signal is rather weak. Moreover, the
differential cross section could be difficult to analyze, since
there are too many parameters involved. In the charged channels the
angular distribution at such a high energy is mostly localized in
the forward direction. This is due to dominance of the pion contact
term, where a virtual pion becomes a real one by the coupling of the
photon with the pion current. Clearly the importance of this
peripheral mechanism makes the angular distribution more difficult
to interpret in terms of low powers of $\cos\theta$.

In the neutral channel $\gamma d\to\pi^0\pi^0np$ the cross section
is more sensitive to resonance contributions. However, the
corresponding angular distribution turns out to be symmetric with
respect to $\cos\theta=0$ regardless of the parity of the resonances
which contribute. This is a trivial consequence of the identity of
the produced pions and thus the corresponding analysis can hardly be
useful. Furthermore, according to the results of Ref.~\cite{2pi}, in
the region of 1 GeV the important role might be played by the
excitation of $F_{15}(1680)$, so that $P_{11}$ or $D_{15}$ seem to
be much less important. In $K^0$ photoproduction the peak although
being more visible is located just on the top of the slope and
therefore could be difficult for experimental identification.

We recall that the calculation presented in Fig.~\ref{fig5} is
performed under the assumption that $P_{11}(1670)$ decays with
almost equal probabilities into all three channels $\eta N$,
$\pi\Delta$, and $K\Lambda$ (see Table~\ref{ta3}). Therefore, if the
partial widths will be better known in the future, the height of the
peaks in Fig.~\ref{fig5} as well as in Fig.~\ref{fig2} could easily
be scaled using, e.g., the formula (\ref{80}). However, we can
expect that our qualitative conclusion about the preference of the
$\eta N$ channel as a tool to study a $P_{11}$ resonance will hardly
be changed, until its mode remains on a level of more than 10-20
$\%$.

\section{Conclusion}

We have calculated photoproduction of $\eta$ mesons in the region
above the $S_{11}$(1535) resonance, where an unusually narrow
structure has been observed in recent experiments for $\gamma
n\to\eta n$ on quasifree neutrons in a deuteron
\cite{Slava,Jaegle:2005,Krusche:2006,kas}. One of our main
objectives was to investigate the impact of Fermi motion and FSI on
the elementary process $\gamma N\to\eta N$. From the presented
studies we are confident that the description within the spectator
model is good in the entire region overlapping $P_{11}(1670)$. As it
is shown the corrections from the FSI are on the percent level so
that the parameters of the single nucleon cross section can be
deduced directly from the data. The Fermi motion effect results in
essential smearing of a narrow structure in the cross section, like
peaks and shoulders, what in some important cases leads to visible
changes in the reaction characteristics. Altogether, we find that
the presence of an exotic narrow resonance with a width about 10-30
MeV can explain the CB-ELSA data \cite{Jaegle:2005,Krusche:2006}.

In a recent paper of~\cite{Shklyar} another mechanism based on the
strong contribution of $P_{11}(1710)$ to $\gamma n\to\eta n$ was
considered. The authors showed that the bump structure in the cross
section can be explained in terms of $S_{11}(1650)+P_{11}(1710)$
contributions without resorting to an exotic narrow $P_{11}(1670)$
state considered in the present paper. In this respect we note that
according to the PDG analysis~\cite{PDG} as well as to the parameter
set presented in~\cite{Shklyar} both $S_{11}(1650)$ and
$P_{11}(1710)$ are coupled more strongly to $\gamma p$ than to
$\gamma n$. Therefore, it is reasonable to expect that the same
mechanism will produce even more pronounced bump in the cross
section for $\gamma p\to\eta p$ which is however not experimentally
observed. In the present paper we suggested another explanation.
Namely, our purpose was to assign the structure to a narrow $P_{11}$
state, which (i) is in accord with the chiral-soliton
model~\cite{DPP}, (ii) has a strong photocoupling to a neutron, and
(iii) could elude identification in PWA due to its small width.

Another point to discuss is whether the available data reveal a true
resonance in the reaction on a neutron at $E_\gamma$=1020 MeV. As is
seen from Fig.~\ref{fig2} crucial is the behavior of the cross
section at 930 MeV, where according to the CB-ELSA results there is
clear minimum. If we assume for the moment, that the minimum
vanishes in the improved data analysis, then we will see only a wide
structure. It might appear as a resonance in the $\sigma_n$ to
$\sigma_p$ ratio simply because of slight minimum exhibited by the
proton cross section at $E_\gamma=1020$ MeV. Therefore, more general
question concerning the experimental results for $\gamma n\to\eta n$
is why the neutron cross section is so high above the $S_{11}(1535)$
region, as compared to what we see on the proton. It is quite
unlikely that here we deal with the interference of the resonances
with the nonresonant background. The latter seems to be
insignificant in $\eta$ photoproduction, primarily due to the
smallness of the $\eta NN$ coupling \cite{Tiator:1994et}. Among the
known resonances there are (according to PDG) no states
predominantly excited on a neutron and, at the same time, having a
sufficiently large $\eta N$ branching ratio. Therefore, even if
future investigations do not support the existence of a narrow
$P_{11}(1670)$, the question about the nature of the $\eta$
photoproduction on the neutron (and eventually on the proton) above
the $S_{11}$ will remain open.

As is noted in~\cite{VPI2} it is a small width that can be used as a
sign that the resonance particle, observed e.g. in meson production
should be attached to the $\overline{10}$ group of baryons and not
to the ordinary baryon octet (in the case of $\overline{10}\otimes
8$ mixture we can speak about dominating antidecuplet part in the
particle wave functions). However it should be noted that if, for
example, the analysis of the partial waves points to the dominance
of an $S_{11}+P_{11}$ configuration in the $\gamma n\to\eta n$
amplitude it could be quite difficult to distinguish between the
'usual' $P_{11}$ with its typical hadronic width of 100 MeV and an
exotic narrow state $P_{11}$ from $\overline{10}$. As the
calculation shows the Fermi motion strongly smears the narrow
resonance structure, so that in both cases we see a broad peak.
Therefore, it is to expect that no simple test exists for
discrimination between the 'ordinary' and exotic $P_{11}$
resonances.

\section*{Acknowledgment}
We thank V.\,Kuznetsov for providing us with GRAAL data, B. Krusche
and I. Jaegle for the preliminary CB-ELSA data and I.\,Strakovsky
for many helpful discussions. The work was supported by the Deutsche
Forschungsgemeinschaft (SFB 443). M.V.P.\ is supported by the Sofja
Kowalewskaja Programme of the Alexander von Humboldt Foundation.

\end{document}